\documentclass[preprint,showpacs,preprintnumbers,amsmath,amssymb,nobibnotes,nofootinbib]{revtex4}
\usepackage{graphicx} 
\usepackage{bm}

\begin{document}

\title{Diffusion coefficients in leaflets of bilayer membranes}
\author{
Kazuhiko Seki}
\affiliation{NRI, 
National Institute of Advanced Industrial Science and Technology (AIST) \\
AIST Central 5, Higashi 1-1-1,Tsukuba, Ibaraki 305-8565, Japan, 
}
\author{Saurabh Mogre
}
\affiliation{Indian Institute of Technology Bombay, Powai, Mumbai, Maharashtra  400 076, India 
}
\author{Shigeyuki Komura 
}
\affiliation{Department of Chemistry, 
Graduate School of Science and Engineering, 
Tokyo Metropolitan University, 
Tokyo 192-0397, Japan
}
\preprint{}
\begin{abstract}
We study diffusion coefficients of liquid domains 
by explicitly taking into account the two-layered structure called leaflets of the bilayer membrane. 
In general, the velocity fields associated with each leaflet are different and 
the layers sliding past each other cause frictional coupling. 
We obtain analytical results of diffusion coefficients for a circular liquid domain in a leaflet,  
and quantitatively study their dependence on the inter-leaflet friction. 
We also show that 
the diffusion coefficients diverge in the absence of coupling between the bilayer and solvents, 
even when the inter-leaflet friction is taken into account. 
In order to corroborate our theory,  the effect of the inter-leaflet friction on the correlated diffusion is examined.

\end{abstract}

\maketitle
\newpage


\section{Introduction}
Diffusion of inclusions in membranes occurs in 2-dimensional (2D) 
 media embedded in solvents, and is different from that in a homogeneous 3-dimensional (3D) fluid. 
In the pioneering work by Saffmann and Delbr\"{u}ck~\cite{saffmandelbruck1975,saffman1976}, 
the membrane was regarded as a thin plane sheet bounded on both sides by solvents. 
Solvents are dragged by the flow in the membrane through the coupling 
to the solvent flow at the 
solvent/membrane interfaces. 
The transfer of momentum between the membrane and solvents necessitates the introduction of a length called 
the Saffmann-Delbr\"{u}ck (SD) hydrodynamic screening length~\cite{saffmandelbruck1975,saffman1976,Oppenheimer}.
The SD expression of the diffusion coefficient has been elaborated to apply to the larger size of objects~\cite{Hughes}. 
Although some experiments were not explained by the SD theory \cite{Gambin}, 
the SD theory and its extended expressions 
have been supported by certain experiments~\cite{Cicuta07,Petrov,Ramadurai09}.

When the membrane is supported on a solid substrate, a very thin layer of solvent exists between 
the membrane and the solid support~\cite{ES88}. 
For supported membranes, the assumption of infinite thickness of solvents in SD theory is irrelevant;  
the diffusion coefficient decreases when the thickness of solvent layer is decreased~\cite{ES88,Stone}.  
The length scale of momentum transfer is characterized by a new length scale, called 
Evans-Sackmann (ES) hydrodynamic screening length in the limit of thin layer of solvent. 
When the size of the diffusing object exceeds the ES hydrodynamic screening length, 
the diffusion coefficient is predicted to show the inverse square size-dependence~\cite{ES88}. 
The inverse square size-dependence has been observed in supported lipid bilayers~\cite{Harb}. 

In the earlier works, diffusion of solid molecules in homogeneous membranes has been studied. 
Recently,  microheterogeneity of the membrane has attracted great interest~\cite{Simons,Allender,Collins,Lingwood}. 
The micro-domains in multicomponent membranes have been expected to be involved in 
signal transduction and control intracellular transport~\cite{Simons,Allender,Collins,Lingwood}. 
By phase separation of 
ternary mixtures of cholesterol, saturated and unsaturated lipids, 
regions rich in saturated lipids
and cholesterol coexist with those rich in unsaturated lipids.    
The smaller regions form circular viscous domains and diffuse laterally. 
The diffusion coefficients of liquid domains have been obtained using 
hydrodynamic calculations by ignoring a shear flow across the membrane~\cite{DeKoker,Ramachandran10,sekiramakomura2011,Fujitani12,Fujitani13}.
However, it is possible to form membranes with bilayer consisting of two leaflets having different compositions~\cite{Allender,Collins}.
The liquid domains in each leaflet do not necessarily lie on top of one another;  
the liquid domains can be formed in one of the leaflets without overlapping to those in the other leaflet. 
In general, the velocity fields associated with each leaflet are different and 
the layers sliding past each other cause frictional coupling~\cite{levinemackintosh2002,seifertlanger1993,stonemcconell1995,camleybrown2013,Han2013}.

When relative motion of the leaflets
does not play a significant role, the bilayer can be considered as a single fluid medium.
However, the inter-leaflet friction would induce different transport in two leaflets of the bilayer for supported membranes,  
because each leaflet faces different environment~\cite{camleybrown2013}.
Furthermore, when the solvent thickness is small, typical for supported membranes, 
the relative velocity difference between the wall and the membrane
induces a shear flow in the solvent through the stick boundary condition. 
The resultant drag force from the solvent can be 
sufficiently large to decouple the two leaflets.

The diffusion of solid particles in a leaflet under the  inter-leaflet friction was  previously 
studied numerically~\cite{camleybrown2013}. 
The complex interplay
between the inter-leaflet friction and the drag from solvents was shown. 
In this paper, we derive analytical results of diffusion coefficients for circular liquid domains. 
Although we use the same mobility tensor derived by Brown {\it et al.}~\cite{camleybrown2013},
we present the analytical expressions showing the effect  of  the  inter-leaflet friction on 
the diffusion coefficients   
for the general thickness of solvent. 
By examining the asymmetry in diffusion coefficients associated with the upper and lower leaflet 
for supported membranes,  
we find that the value of inter-leaflet friction can be estimated. 
In addition, we show that the momentum is dissipated by the inter-leaflet friction, but 
this dissipation mechanism is not sufficient to overcome logarithmic 
long-range correlations in the flow fields leading to a divergence of the diffusion coefficient. 
In order to corroborate our theory,  the correlated diffusion is studied 
under the influence of the inter-leaflet friction.  

In experiments, the frictional interaction between the leaflets
has been studied in detail by applying the interlayer shear stress induced by a large local deformation~\cite{Evans94,Pott,Raphael,Bitbol}. 
Zhang and Granick concluded that the difference in diffusion coefficient of lipids between two leaflets is small 
for supported bilayers~\cite{Zhang}.
Detailed investigation on diffusion of lipids in leaflets on supported membranes suggests that  
the diffusion coefficient of lipids in the lower leaflet is reduced 
from that of the upper leaflet depending on the type of substrate~\cite{Scomparin,Hetzer,Mechan}.

Simulations have been also performed to 
estimate the value of inter-leaflet friction~\cite{Shkulipa,denOtter,Xing}. 
From coarse-grained molecular simulations, 
the lateral diffusion of lipids in the lower leaflet to the solid substrate 
was shown to be slower than that of the upper leaflet  by one order of magnitude~\cite{Xing}.
We discuss the results on the asymmetry in diffusion coefficients for the supported membranes 
using the values 
of the inter-leaflet friction thus estimated. 

In Sec.~\ref{sec:hydrodynamics}, the diffusion coefficient of a circular domain in a bilayer membrane is obtained by taking into account 
the inter-leaflet friction for arbitrary thickness and viscosity of solvents. 
In Sec.~\ref{sec:symmetricenvironment}, 
we study the effect  of the inter-leaflet friction on the SD length and ES length. 
Symmetric environments are considered for simplicity. 
Diffusion in a supported bilayer is considered in Sec.~\ref{sec:supportedbly}.
In Sec.~\ref{sec:CorrelatedDynamics}, we study correlated dynamics of two point particles. 
In Sec.~\ref{sec:Discussion}, we show that the finite diffusion coefficient cannot result from 
the inter-leaflet friction alone without solvents.  
The final section is devoted for summary and discussion.

\section{The diffusion coefficient of a circular domain in a leaflet}
\label{sec:hydrodynamics} 
\subsection{Hydrodynamic model}
\begin{figure}[ht!]
\centering
\includegraphics[width=0.5\textwidth]{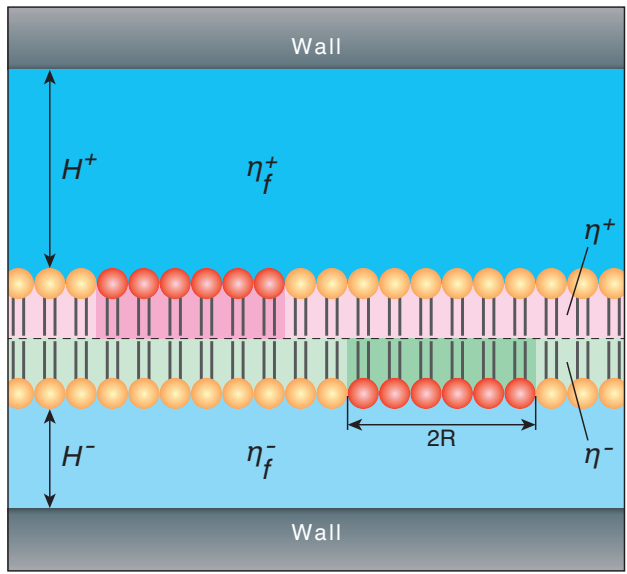}
\caption{(Color online)  The geometry of the bilayer in a general environment. Liquid domains of size $2R$ diffuse in leaflets.}
\label{fig:system}
\end{figure}

We explicitly take into account the two-layered structure of the membrane as shown in Fig.~\ref{fig:system}. 
The structure is sometimes referred to as the leaflets 
and its hydrodynamics has been studied previously \cite{camleybrown2013,seifertlanger1993}. 
These leaflets are denoted by the indices ``$+$'' and ``$-$'' for the upper and lower monolayer, respectively. 
The monolayers of the membrane, called leaflets, are coupled through the drag force between the upper and lower leaflets. 

We consider a circular liquid domain of radius $R$ which is immersed in either leaflet. 
For simplicity, we assume that the viscosity of the liquid inside the domain is the same as that of the leaflet in which the domain is located.
We also assume that the domain maintains a circular shape by the line tension and does not undergo deformation as it moves. 
The system is illustrated in Fig.~\ref{fig:system}. 
We will derive the general expression for the diffusion coefficient in either case. 

The 2D viscosity of the leaflets are denoted by $\eta^\pm$, depending on the upper or lower leaflet. 
The bilayer membrane is sustained  in a solvent 
at distance $H^+$ and $H^-$ below and above the walls, respectively. 
The viscosity of the solvent above and below the membrane is represented by $\eta_f^+$ and $\eta_f^-$, respectively.

The in-plane flow in each leaflet can be described using Stokes equations with additional terms associated with   
the coupling between the leaflet and the solvent through the interface and the friction between leaflets~\cite{levinemackintosh2002,seifertlanger1993,stonemcconell1995,camleybrown2013},
\begin{equation}
\eta^\pm\nabla^2\bm{v}^\pm-\nabla p^\pm +K^\pm*\bm{v}^\pm \mp \Lambda(\bm{v}^+-\bm{v}^-)+\bm{f}^\pm=0, 
\label{eq:ns1}
\end{equation}
with the incompressibility condition, 
\begin{equation}
\nabla \cdot \bm{v}^\pm=0, 
\label{eq:ns2}
\end{equation}
where $p^\pm$ and $\bm{v}^\pm$ indicate the hydrodynamic pressure and fluid velocities of the leaflets, respectively. 
In the above, the position vector ${\bm r}$ is abbreviated. 
The term $\Lambda (\bm{v}^+-\bm{v}^-)$ accounts for the frictional coupling between the two leaflets. 
The convolution term 
$K^\pm*\bm{v}^\pm=\int d^2r^{'} K^\pm ( \bm{r} - \bm{r^{'}} )\bm{v}^\pm(\bm{r^{'}})$ represents 
the drag from the solvent.
Using Fourier transform $\hat{K}(\bm{k})=\int d^2r \exp \left( - i \bm{k} \cdot \bm{r} \right)K(\bm{r})$, 
the convolution can be expressed as a product. 
The drag force from the solvent depends on the geometry of the system. 
For the system under consideration, we have $\hat{K}^\pm=-\eta_f^{\pm}k\coth(kH^\pm)$ for arbitrary solvent thickness~\cite{lubenskygoldstein1996,inaurafujitani2008,fischer2004,Ramachandran11}. 
The external in-plane force is denoted by $\bm{f}^\pm$. 

\subsection{Diffusion coefficient}
We consider the case when the circular domain moves with a velocity ${\bm U}$ in leaflet $\alpha$. 
The origin of the coordinates is taken to be the center of the domain and the $x$-coordinate is chosen in the direction of ${\bm U}$.
The external in-plane force 
applied at the periphery of the circular domain in Eq.~(\ref{eq:ns1}) is given by  
$\bm{f}^\pm= f^{(\ell {\rm n})} \cos \theta \, {\bm n} \delta (r-R)/(2 \pi R)$ \cite{DeKoker}, 
where ${\bm n}$ is the outward normal unit vector on the circumference of the domain, $\theta$ is the angle between ${\bm U}$ and ${\bm r}$, and 
$\cos \theta$ term reflects the symmetry of the system. 
$f^{(\ell {\rm n})}$ is a constant which should give the external force after the integration over the circular domain boundary. 

We can derive the diffusion coefficient of the circular domain in either leaflet 
using a similar method as in our previous publication~\cite{sekiramakomura2011}. 
By introducing a component of a mobility tensor in Fourier space given in the Appendix A \cite{camleybrown2013},
the diffusion coefficient of the liquid domain is obtained as,   
\begin{equation}
D_{dom}^{\alpha}=k_{\rm B}T\int_0^\infty{\rm d}k\frac{\left[J_1(kR)\right]^2}{\pi kR^2}
\left[\frac{\eta^{(-\alpha)}k^2+
\eta_f^{(-\alpha)}k\coth(kH^{(-\alpha)})+\Lambda}{\prod_{\alpha} (\eta^{(\alpha)}k^2+\eta_f^{(\alpha)}k\coth(kH^{(\alpha)})+\Lambda)-\Lambda^2}\right] ,
\label{eq:dgen}
\end{equation}
where $\alpha$ denotes the leaflet, $-\alpha$ denotes the leaflet opposite to $\alpha$ 
and $J_1(z)$ denotes the Bessel functions of the first kind of order $1$~\cite{abram-stegun}.

The expression for the diffusion coefficient contains various physical quantities with varying magnitudes and different units. 
In order to systematically compare various cases and to have a concise understanding of the results,  
we make the variables dimensionless mainly using 
the SD hydrodynamic screening length given by 
$1/\nu^\pm\equiv\eta^\pm/\eta_f^\pm$  
~\cite{saffmandelbruck1975,saffman1976,Oppenheimer}. 
The following dimensionless variables are defined:  
\begin{equation}
\rho^\pm=\nu^{\pm}R, 
\quad~h^\pm=\nu^{\pm}H^\pm, 
\quad~\lambda^\pm=\frac{\Lambda}{\eta^\pm}\frac{1}{\left(\nu^{\pm} \right)^2},
\quad~\mu=\frac{\eta^-}{\eta^+}.
\label{dim}
\end{equation}
Here, $\rho^\pm$ and $h^\pm$ represent the dimensionless radius of the liquid domain and the dimensionless solvent thickness, respectively. 
The ES hydrodynamic screening length is given by the smaller of $\sqrt{h^\pm}/\nu$. 
 A new length, $\sqrt{\eta^-/\Lambda}$, is associated with the inter-leaflet friction. 
 This quantity can be called the inter-leaflet sliding length, and 
 contributes to the hydrodynamic screening factor in addition to the contribution from the solvent. 
 The variable $\mu$ is the ratio between the 3D-viscosities of the leaflets. 
In the later sections, the index of the dimensionless variables is dropped when it is redundant, {\it e.g.} 
$h=h^\pm$ whenever it is clear from the context or the variables do not differ between the indices.

\section{Diffusion for a liquid domain in a symmetric environment}
\label{sec:symmetricenvironment} 
\begin{figure}[ht!]
\centering
\includegraphics[width=0.5\textwidth]{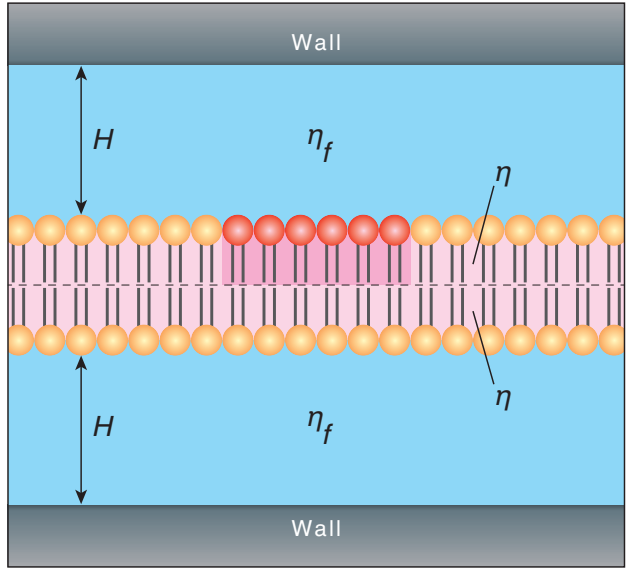}
\caption{\label{fig:sym}(Color online) The geometry of the bilayer in a symmetric environment. Liquid domains of size $2R$ 
shown by red diffuse in leaflets.}
\end{figure}

For simplicity we consider 
the symmetric system as illustrated in Fig.~\ref{fig:sym}. 
We assume a solvent layer of equal thickness on both sides of the membrane, i.e. $H^+=H^-=H$. 
The viscosity of the solvents and two leaflets are also taken to be equal, i.e. $\eta_f^+ = \eta_f^- = \eta_f$ and $\eta^+ = \eta^- = \eta$.

For this case, the diffusion coefficient is simplified to, 
\begin{equation}
D_{sym}=\frac{1}{2}\left[D_0+D_1 (\lambda) \right],
\label{eq:drel}
\end{equation}
where
\begin{equation}
D_1 (\lambda)=\frac{k_{\rm B}T}{\pi\eta\rho^2}\int_0^\infty {\rm d}\kappa\frac{\left[J_1(\kappa\rho)\right]^2}{\kappa}\left[\frac{1}{\kappa^2+\kappa\coth(\kappa h)+2\lambda}\right]. 
\label{eq:dsym}
\end{equation}
In the above, $D_0$ represents 
$D_1(\lambda=0)$, which is the case in the absence of the inter-leaflet friction. \cite{sekiramakomura2011}

\begin{figure}[h]
\includegraphics[width=0.5\textwidth]{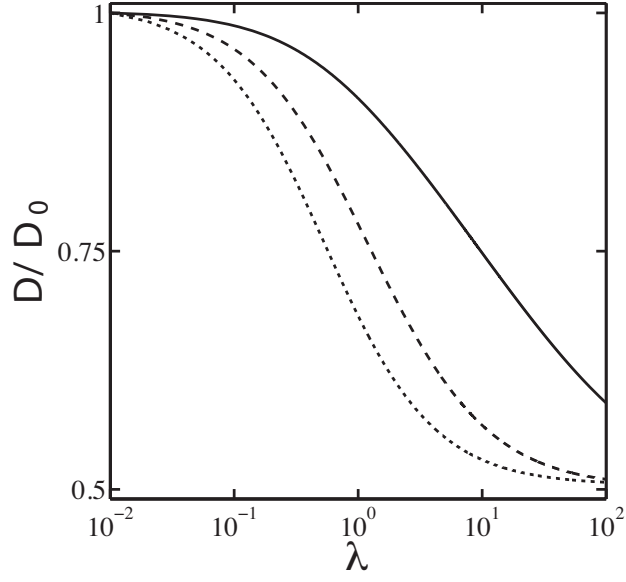} 
\caption{\label{DsymvsBeta} Normalized diffusion coefficient $D/D_0$ as a function of $\lambda$ for $ h=1$ is shown by using Eq.~(\ref{eq:drel}). Here, $D_0=D(\lambda=0)$. The values of $\pi \eta D_0/\left(k_{\rm B} T\right)$ are $0.5882(\rho=0.1)$, $0.1404(\rho=1)$, and $0.0044(\rho=10)$. 
The solid, dashed and dotted lines represent $\rho=0.1,1,10$, respectively.} 
\end{figure}

We can see from Eq.~(\ref{eq:drel}) and Fig. \ref{DsymvsBeta} that the diffusion coefficient decreases from $D_0$ to $D_0/2$ 
as the inter-leaflet friction is changed from 0 to $\infty$. 
The qualitatively similar behavior is obtained by Camley and Brown for a solid circular object in a freely suspended membrane when 
the symmetric leaflets are faced to the same solvent medium  \cite{camleybrown2013}. 
We also note  in Fig. \ref{DsymvsBeta} 
that the diffusion coefficient as a function of $\lambda$ behaves very differently when $\rho \ll 1$. 
The reason is separately discussed in Appendix C. 

\subsection{Infinite solvent medium}
In the limit of $H \to \infty$, we have $\coth(\kappa h)\approx1$ in Eq.~(\ref{eq:dsym}).  
The expression for the diffusion coefficient becomes  
\begin{equation}
D_{sym}=\frac{k_{\rm B}T}{2\pi\eta\rho^2}\int_0^\infty {\rm d}\kappa\frac{\left[J_1(\kappa\rho)\right]^2}{\kappa}\left[ \frac{1}{\kappa^2+\kappa}+\frac{1}{\kappa^2+\kappa+2\lambda} \right]. 
\label{hinf}
\end{equation}
The above integral can be calculated  by using Mathematica 
and is expressed in terms of Meijer-G functions~\cite{mathematica,note2}, 
\begin{multline} 
D_{sym}=\frac{k_{\rm B}T}{4\pi\eta\rho^2}
\left[ \frac{1}{\sqrt{1-8\lambda}}
    \left\{  
    \frac{1}{\pi^{3/2}c_1}G_{2~4}^{3~2}
        \left( (c_1\rho)^2
            \left| \begin{array}{c}
                    \frac{1}{2},\frac{1}{2} \\
                    0,\frac{1}{2},1,-1 \\
                    \end{array}
            \right.
        \right) 
        \right. \right.\\ \left.
         -\frac{1}{\pi^{3/2}c_2}G_{2~4}^{3~2}
        \left( (c_2\rho)^2
            \left| \begin{array}{c}
                    \frac{1}{2},\frac{1}{2} \\
                    0,\frac{1}{2},1,-1 \\ 
                    \end{array}
            \right.
        \right)
    \right\} \\ \left.
+\frac{1}{2\lambda} -1-\frac{1}{\pi^{3/2}}G_{2~4}^{3~2}
        \left( \rho^2
            \left|
            \begin{array}{c} 
            \frac{1}{2},\frac{3}{2} \\
            0,1,\frac{3}{2},-1 \\
            \end{array}
            \right.
        \right)
\right], 
\label{hinfanal}
\end{multline}
where $c_1$ and $c_2$ are the roots of the quadratic equation, $\kappa^2+ \kappa + 2\lambda=0$.
The limits for the domain size can now be taken to this expression.

By taking the limit of small domain size, i.e., $\rho\ll 1$, $c_1 \rho\ll 1$, and $c_2 \rho\ll 1$, 
in Eq.~(\ref{hinfanal}), 
the diffusion coefficient is shown to depend logarithmically on 
the domain size, 
\begin{equation}
D_{sym}\approx\frac{k_{\rm B} T}{4\pi \eta}\left[\ln \left(\frac{2 \eta}{\eta_f R}\right) -\gamma + \frac{1}{4}+\frac{1}{2(c_1-c_2)}\left(c_2 \ln c_2 -
c_1 \ln c_1 \right)\right] ,  
\label{eq:asym_infinite}
\end{equation}
where $\gamma= 0.5772 \cdots $ is Euler's constant~\cite{abram-stegun}.

By taking the limit of large domains, i.e. $c_1 \rho \gg 1$, $\rho \gg 1$ and $c_2 \rho \gg 1$, in Eq.~(\ref{hinfanal}), we obtain, 
\begin{equation}
D_{sym}\approx \frac{2k_{\rm B} T}{3\pi^2 \eta_f R} + \frac{k_{\rm B} T}{8\pi R^2}\left(\frac{1}{\Lambda}-\frac{2\eta}{\eta_f^2}\right) .
\label{eq:asyminfinity}
\end{equation}
In the asymptotic limit of large domain size, the diffusion coefficient  is independent of the membrane viscosity $\eta$. 
The asymptotic domain size-dependence is given by the first term, $1/R$, which is independent of the inter-leaflet friction. 
The second term showing $1/R^2$-dependence is associated with the inter-leaflet friction.  

Equation (\ref{eq:asyminfinity}) implies that 
the diffusion coefficient   depends on $\lambda$ if it is smaller than $\lambda^*_{sym} = 3 \pi/(16 \rho)$. 
By plotting the diffusion coefficient as a function of $\lambda$ as shown in Fig. \ref{DsymvsBeta}, 
we confirm that the inflection point where $D_{sym}$ depends most strongly on $\lambda$,  
can be estimated by using $\lambda^*_{sym}$ when $\rho \geq 0.1$.

\subsection{Thin solvent medium}
In the limit of, $H \to 0$, we have $\coth(\kappa h)\approx 1/\kappa h$ in Eq.~(\ref{eq:dsym}).  
By taking this limit, the expression for the diffusion coefficient  becomes  
\begin{equation}
D_{sym}=\frac{k_{\rm B}T}{2\pi\eta\rho^2}\int_0^\infty {\rm d}\kappa
\frac{\left[J_1(\kappa\rho)\right]^2}{\kappa}\left[\frac{1}{\kappa^2+1/ h+2\lambda} + \frac{1}{\kappa^2+1/ h}\right]. 
\label{eq:hzero}
\end{equation}
We can analytically integrate Eq.~(\ref{eq:hzero}) and the result is given by,  
\begin{multline}
D_{sym}=\frac{k_{\rm B}T}{2\pi R^2}
\left\{\frac{H}{\eta_f}\left[\frac{1}{2}-I_1\left(R\sqrt{\frac{\eta_f}{\eta H}}\right)K_1\left(R\sqrt{\frac{\eta_f}{\eta H}}\right)\right]\right.\\\left.+
\frac{1}{(\eta_f/H)+2 \Lambda}\left[\frac{1}{2}-I_1\left(R\sqrt{\frac{(\eta_f/H)+2 \Lambda}{\eta}}\right)K_1\left(R\sqrt{\frac{(\eta_f/H)+2 \Lambda}{\eta}}\right)\right]\right\} .
\label{eq:h0anal}
\end{multline}

By taking the limit of $\rho \ll 1$, $\rho/\sqrt{ h} \ll 1$ and $\rho/\sqrt{(1/ h)+2\lambda} \ll 1$ in Eq.~(\ref{eq:h0anal}), we obtain, 
\begin{equation}
D_{sym}\approx 
\frac{k_{\rm B} T}{4\pi\eta}\left[\ln\left(\frac{2}{R} 
\sqrt{\frac{\eta_f H}{\eta}}
\frac{1}{\left[1+(2\Lambda H/\eta_f)\right]^{1/4} }
\right)-\gamma+\frac{1}{4} \right].
\end{equation}
As in the case of an infinite solvent medium, Eq.~(\ref{eq:asym_infinite}), 
the diffusion coefficient is shown to depend logarithmically on the domain size. 

By taking the limit of $\rho \gg 1$, $\rho/\sqrt{ h} \gg 1$ and $\rho/\sqrt{(1/ h)+2\lambda} \gg 1$, 
Eq.~(\ref{eq:h0anal}) reduces to,  
\begin{equation}
D_{sym} \approx \frac{k_{\rm B} T H}{2\pi \eta_f R^2}\left[\frac{1+\Lambda H/\eta_f}{1+2\Lambda H/\eta_f}\right]. 
\label{eq:Dsym_asympt}
\end{equation}
As in the case of an infinite solvent medium, 
the diffusion coefficient is independent of the membrane viscosity.  
The diffusion coefficient shows $1/R^2$-dependence in this limit, 
which is stronger than that in Eq. (\ref{eq:asyminfinity}).  
The size dependence of the diffusion coefficient is qualitatively the same as that obtained by regarding 
the membrane as a sheet of 2D fluid without any leaflet structure, though 
there are quantitative differences given by $\Lambda H/\eta_f$-dependence. 
\subsection{Arbitrary thickness of solvent medium}

The integrand in Eq.~(\ref{eq:dsym}) can be expressed by a sum using the partial fraction expansion as shown in Appendix B. 
The result after the integration is given as  
\begin{equation}
D_{sym}=\frac{k_{\rm B}T}{2\pi\eta \rho^2} \sum_{s=1}^{\infty} \left\{W_s' \left[ \frac{1}{2}-I_1(\omega_s'\rho) K_1(\omega_s'\rho)\right] + W_s \left[ \frac{1}{2}-I_1(\omega_s\rho) K_1(\omega_s\rho)\right]\right\}, 
\label{eq:pfsym}
\end{equation}
where the weight factor is given by
\begin{equation}
W_s=\frac{2}{ h\omega_s^4+(1+ h-4\lambda h)\omega_s^2+2\lambda+4 h\lambda^2}
\end{equation}
and $\omega_s$ satisfies the characteristic equation,
\begin{equation}
\omega_s^2=\omega_s\text{cot}(\omega_s h)+2\lambda. 
\label{eq:characteristiceq}
\end{equation}
In the above, $W_s'=2/[\omega_s'^{2}(1 +  h+\omega_s'^{2} h)]$ and $\omega'_s=\text{cot}(\omega'_s h)$ are $W_s$ and $\omega_s$ 
obtained by setting $\lambda=0$, respectively. 
Moreover, $I_1(z)$ and $K_1(z)$ are the modified Bessel functions of order 1~\cite{abram-stegun}. 
We can show that 
the rigorous condition for $H \rightarrow 0$ to obtain Eq. (\ref{eq:h0anal}) is $ h=\nu H<\pi$. 
This is the condition to obtain Eq.~(\ref{eq:h0anal})  from Eq.~(\ref{eq:pfsym}) 
by retaining the first term in the expansion and ignore the rest.

In the above, $1/(\omega_s \nu)$ represents the hydrodynamic screening length under the inter-leaflet friction. 
The diffusion coefficient is given by the infinite sum of the expression associated with different hydrodynamic screening lengths. 
In this sense, hydrodynamic screening is related to multiple length scales given by $1/(\omega_s \nu)$ with $s=1,2,3, \cdots$. 
The weight approximately satisfies the relation $W_s \sim 1/\omega_s^4$. 
In the below, we assume that $1/(\omega_s \nu)$ is sorted from small to large values. 
The difference between the successive screening length, $1/(\omega_s \nu)$ and $1/(\omega_{s+1} \nu)$, 
is approximately equal to $\pi/ h$. 
The weight factor $W_s$ rapidly decreases if $ h=\nu H<\pi$. 
In this case, $1/(\omega_1 \nu)$ can be regarded as the virtual hydrodynamic screening length.
The difference due to retaining only the first term in the diffusion coefficient 
is numerically investigated, and it is at most 10\% when $ h=1$. 
The difference increases by increasing the value of $ h$.

\begin{figure}[h]
\includegraphics[width=0.5\textwidth]{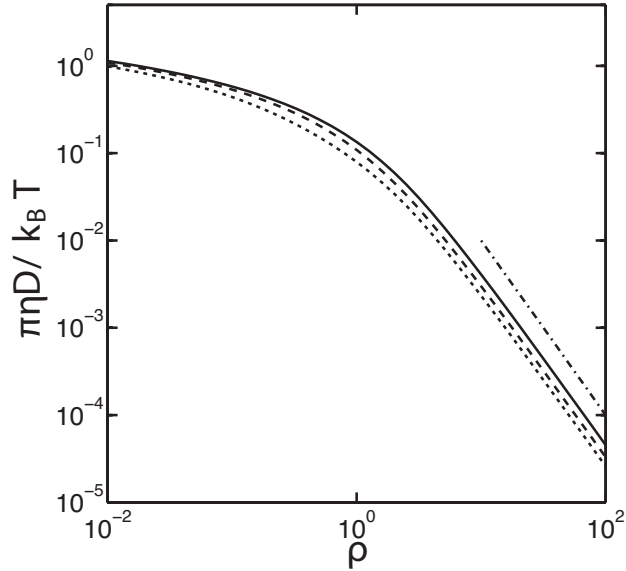} 
\caption{\label{Dsymvsrho} $\pi \eta D/k_{\rm B} T$ as a function of $\rho$  for $ h=1$ is shown by using Eq.~(\ref{eq:drel}).  The solid, dashed and dotted lines represent $\lambda=0.1,1,10$, respectively. 
The dash-dot line indicates the asymptotic $1/\rho^2$ dependence. } 
\end{figure}

In Fig. \ref{Dsymvsrho} 
we show the diffusion coefficient as a function of the domain size. 
The diffusion coefficient decreases by increasing $\lambda$. 
When $\rho$ is small, 
the variation of the diffusion coefficients for the different values of the inter-leaflet friction decreases.  
This can be understood by noticing that 
the diffusion coefficient depends logarithmically on the strength of the inter-leaflet friction. 

\section{Diffusion in a supported bilayer}
\label{sec:supportedbly}
\begin{figure}[h]
\centering
\includegraphics[width=0.5\textwidth]{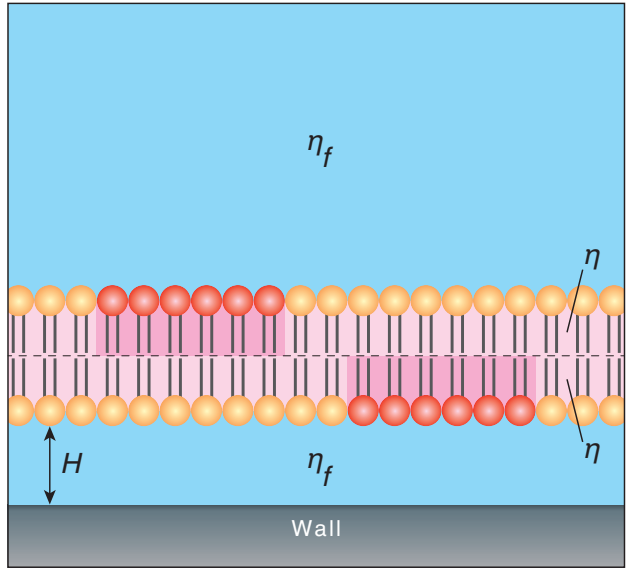}
\caption{\label{fig:asym}(Color online)  The geometry of the bilayer in an asymmetric environment. 
Liquid domains of size $2R$ shown by red diffuse in leaflets.}
\end{figure}

In this section, we consider the case where asymmetry in the system is introduced by changing 
the thickness of the solvent layer 
above and below the membranes. 
The values of the viscosities are kept the same as in the symmetric case for simplicity, 
although a similar calculation is possible with different values as well. 
Figure \ref{fig:asym} illustrates the system being considered. 

\subsection{Formally exact results}
\label{subsec:Formal}
We note that the domain under consideration can be located in either of the two leaflets and 
diffuses differently in each case. 
We denote the diffusion coefficient  when the domain is in the upper and lower leaflets by $D^+$ and $D^-$, respectively. 

The situation of supported bilayer can be mathematically expressed as 
$H^+ \to \infty$ and $\nu^- H^- \ll 1$. 
In this limit, we have $\coth(\kappa h^+)\approx 1$ and $\coth(\kappa h^-)\approx 1/\kappa h^-$.
By taking above limits and using Eq.~(\ref{eq:dgen}), the expression for the diffusion coefficient   becomes, 
\begin{subequations}
\begin{equation}
D_{sup}^{\alpha}=\frac{k_{\rm B} T}{\pi\eta\rho^2}\int_0^{\infty}{\rm d}\kappa\frac{[J_1(\kappa\rho)]^2}{\kappa}\left[\frac{g^{\alpha}(\kappa)}{g^+(\kappa)g^-(\kappa)-\lambda^2}\right] ,
\end{equation}
where $g^{\alpha}$ is given by, 
\begin{equation}
g^+(\kappa)=\kappa^2+\frac{1}{ h}+\lambda, 
\end{equation}
and 
\begin{equation}
g^-(\kappa)=\kappa^2+\kappa + \lambda , 
\end{equation}
\label{Dphys}
\end{subequations}
respectively. 

As before, by applying a partial fraction expansion to Eq.~(\ref{Dphys}),  
we obtain  \cite{mathematica}
\begin{equation}
D_{sup}^{\alpha}=\frac{k_{\rm B} T}{\pi\eta\rho^2}\sum_{i=1}^4\frac{g^{\alpha}(\omega_i)}{f'(\omega_i)}\left(\frac{-1}{2\omega_i}\right)\left[1-
\frac{1}{\pi^{3/2}}G_{2~4}^{3~2}
        \left((\omega_i \rho)^2
            \left|
            \begin{array}{c} 
            \frac{1}{2},\frac{1}{2} \\
            0,\frac{1}{2},1,-1 \\
            \end{array}
            \right.
        \right)
    \right] ,
    \label{eq:phyis_general}
\end{equation}
where $\omega_i$ are the roots of the denominator, $f(\kappa) =g^+(\kappa)g^-(\kappa)-\lambda^2=0$ and their real parts are 
negative,  
while $f'(\kappa)$ is the derivative of the denominator with respect to $\kappa$. 
We note that $\omega_i$ are independent of $\rho$, and use this fact in understanding the size dependence.

As in the previous section, by taking limits of $\rho$ in Eq.~(\ref{eq:phyis_general}), 
we obtain for $\omega_i \rho \ll 1$,
\begin{equation}
D_{sup}^{\alpha}\approx \frac{k_{\rm B} T}{4\pi \eta} \sum_{i=1}^4 \frac{\omega_i g^{\alpha}(\omega_i)}{f'(\omega_i)}
\left[\ln\left(\frac{2}{\omega_i \rho}\right)-\gamma + \frac{1}{4}\right], 
\end{equation}
and 
for $\omega_i \rho \gg 1$
\begin{equation}
D_{sup}^{\alpha}\approx\frac{k_{\rm B} T}{2\pi \eta} 
\sum_{i=1}^4 \frac{g^{\alpha}(\omega_i)}{f'(\omega_i)} \left(\frac{-1}{\omega_i \rho^2}\right). 
\end{equation}
As in the previous case, 
the diffusion coefficient   depends logarithmically on the domain size for small domains, or 
 the inverse square of the domain size  for large domains.

\subsection{Approximate results}
\label{subsec:Formal}
Although we have obtained the formal solution, Eq.~(\ref{eq:phyis_general}), in terms of the 
 roots of a fourth order polynomial equation, 
 finding the $\lambda$ or $ h$ dependence of the roots is a formidable task. 
In order to avoid this difficulty, we simplify the model by taking the limit of $\eta_f^+ \rightarrow 0$. 
When the solvent is water, 
the solvent viscosity is about two order of magnitude smaller than that of lipids. 
The limit of $\eta_f^+ \rightarrow 0$ could be justified for a wide range of the values of $ h$, $\lambda$ and $\rho$. 
We thoroughly investigate the mathematical structure of the diffusion constant in the limit of $\eta_f^+ \rightarrow 0$. 
Then, we compare the analytical solution in the limit of $\eta_f^+ \rightarrow 0$ and that 
with a finite $\eta_f^+ $. 

By taking $\eta_f^+ =0$ in Eq.~(\ref{eq:dgen}) and assuming that 
the solvent layer below the membrane is  very thin, i.e., $\nu^- H^- \ll 1$, 
we get, 
\begin{subequations}
\begin{equation}
D_{sup}^{\alpha}=\frac{k_{\rm B} T}{\pi \eta^- \rho^2}\int_0^{\infty}{\rm d}\kappa \frac{[J_1(\kappa \rho)]^2}{\kappa}
\left[\frac{g^{\alpha}(\kappa)}{g^+(\kappa)g^-(\kappa) -\lambda^- \lambda^+}\right], 
\label{eq:floating}
\end{equation}
where we denote 
\begin{eqnarray}
g^+(k)=\kappa^2 + 1/ h^- +\lambda^-, \mbox{ }
g^-(k)= \kappa^2 +\lambda^+ . 
\end{eqnarray}
\end{subequations}
For simplicity, we consider the case when the viscosity of the upper leaflet is the same as that of the lower leaflet.

\begin{figure}[!ht]
\centering 
\includegraphics[width=0.5\textwidth]{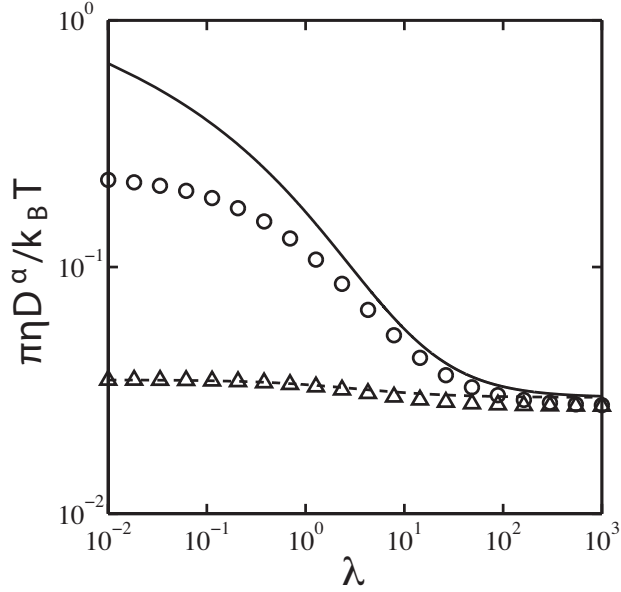} 
\caption{\label{Dphys2vsbeta} $\pi \eta D^{\alpha}/k_{\rm B} T$ as a function of $\lambda$ is shown for $ h=0.1$ and $\rho=1$. 
The circular and triangular markers denote $D^+$ and $D^-$ for the supported case, while, 
the solid and the dashed lines denote $D^+$ and $D^-$ in the limit of $\eta_f^+ \rightarrow 0$, respectively.}
\end{figure}

The integral of Eq.~(\ref{eq:floating}) can be evaluated analytically to give, 
\begin{subequations}
\begin{multline}
D_{sup}^{\alpha}=\frac{k_{\rm B} T}{\pi R^2}\left[\frac{d^\alpha}{2b}+\frac{H}{\eta_f}\frac{1}{ c^+-c^-  } 
\left\{ \left(1-
\frac{d^\alpha}{c^-} \right)I_1\left(R\sqrt{\frac{\nu c^-}{H}}\right)K_1\left(R\sqrt{\frac{\nu c^-}{H}}\right) 
\right. \right.\\ \left. \left. 
-\left(1- \frac{d^\alpha}{c^+} \right)I_1\left(R\sqrt{\frac{\nu c^+}{H}}\right)K_1\left(R\sqrt{\frac{\nu c^+}{H}}\right)\right\}\right] ,
\label{eq:floating3}
\end{multline}
where we have introduced the notations, 
\begin{equation}
d^+=1+ \frac{\Lambda H}{\eta_f}, \quad
d^-=\frac{\Lambda H}{\eta_f},
\end{equation}
and 
\begin{equation}
c^{\pm}=\frac{1}{2} \left[1 + 2 \frac{\Lambda H}{\eta_f} \pm \sqrt{1+ \left(2 \frac{\Lambda H}{\eta_f} \right)^2} \right], 
\end{equation}
\label{eq:dphys2}
\end{subequations}
where the double sign corresponds. 
The first term shows $1/R^2$ size-dependence.  
It is independent of the membrane viscosity, and is a linear function of the inverse of the inter-leaflet friction constant. 
 It can be seen from Eq.~(\ref{eq:dphys2}) that when $1 \approx 2\Lambda H/\eta_f$ ($1/ h \approx 2\lambda$), 
$D$ changes 
rapidly 
 by changing the inter-leaflet friction coefficient. 
The diffusion coefficient as a function of $\lambda$ for a supported bilayer is shown in Fig.~\ref{Dphys2vsbeta}. 
Though the argument is not rigorous, 
we see from Fig.~\ref{Dphys2vsbeta} that the diffusion coefficient indeed changes 
when $\lambda$ is close to the above mentioned value. 
This value is equal to $\lambda^*_2$ obtained for the domain in a symmetric environment 
(see Appendix C for detailed discussion on $\lambda^*_2$). 
By decreasing $\rho$, we note that the inflection point is not characterized by $\lambda^*_2$ but is close to $\lambda^*_1$ as 
in the case of a symmetric environment. (see Appendix C for the detailed discussion on $\lambda^*_1$).

For $R\sqrt{\nu c^\alpha/H} \ll 1$, Eq.~(\ref{eq:floating3}) can be expanded to yield, 
\begin{equation}
D_{sup}^{\alpha}\approx \frac{k_{\rm B} T }{4 \pi \eta}
\left[\ln \left( 
\frac{2}{R} 
\sqrt{\frac{\eta_f H}{\eta}}
\left(\frac{\eta_f}{\Lambda H}\right)^{1/4} 
\right)
- \gamma +\frac{1}{4}
+
\frac{\ln(c^\alpha/c^{(-\alpha)})}{4\sqrt{1+ \left(2 \Lambda H/\eta_f \right)^2}}
\right], 
\end{equation}
where $(-\alpha)$ represents the leaflet opposite to $\alpha$ as before. 
The diffusion coefficient   shows a logarithmic size-dependence in this limit. 
The condition, $R\sqrt{\nu c^\alpha/H} \ll 1$, cannot be satisfied in the $\Lambda \rightarrow \infty$ limit 
because $c^\alpha$ given by 
Eq. (\ref{eq:dphys2}) diverges. 
The order of the limit is important in this respect.

For $R\sqrt{\nu c^\alpha/H} \ll 1$, the diffusion coefficient   simplifies into  
\begin{equation}
D_{sup}^\alpha \approx \frac{k_{\rm B} Td^\alpha}{2\pi R^2 \Lambda}.
\end{equation}
This diffusion coefficient   scales with the domain size as $1/R^2$. 
In this limit, $D_{sup}^-$ is independent of the value of the inter-leaflet friction coefficient. 

As seen in Fig.~\ref{Dphys2vsbeta}, 
the $\lambda$-dependence of the
diffusion coefficient for the supported case is 
well captured by that obtained in the limit of $\eta_f^+ \rightarrow 0$. 
Both of these cases have a large asymmetry between $D^+$ and $D^-$ as the friction between the leaflets changes. 
This implies that contribution to friction from the fluid above the bilayer is small, and 
the friction from the lower fluid is dominant. 
$D^-$ has a very low friction dependence and does not change much by changing $\lambda$. 

The $\lambda$-dependence can be rationalized by taking the limits. 
In the limit of $\lambda \rightarrow \infty$, 
Eq.~(\ref{eq:floating}) becomes, 
\begin{align}
D_{sup}^{\alpha}&=\frac{k_{\rm B} T}{2 \pi \eta \rho^2}\int_0^{\infty}{\rm d}\kappa \frac{\left[J_1(\kappa \rho)\right]^2}{\kappa}
\frac{1}{\kappa^2 + 1/(2 h)}
\\
&= \frac{k_{\rm B} T H}{2 \pi \eta_f R^2} \left( 
\frac{1}{2}- I_1 \left( R \sqrt{\frac{\nu}{2H}} \right)K_1 \left( R \sqrt{\frac{\nu}{2H}} \right)
\right) .
\label{eq:floating1}
\end{align}
Figure~\ref{Dphys2vsbeta} shows that for very large values of $\lambda$, 
$D^+$ and $D^-$ converge to the same value for both the cases. 
This is because the leaflets are moving together and losing their bilayer nature due to the high inter-leaflet friction.
Whereas in the limit of $\lambda \rightarrow 0$, 
$D_{sup}^{+}$ in Eq.~(\ref{eq:floating}) shows a divergence, and  
$D_{sup}^{-}$  in Eq.~(\ref{eq:floating}) reduces to the same expression as 
Eq.~(\ref{eq:floating1}) with $\nu/2$ replaced by $\nu$. 
This replacement implies that the monolayer thickness should be used instead of the bilayer thickness to obtain 
the membrane viscosity in the absence of the inter-leaflet friction. 
Note that the 2D-membrane viscosity is its 3-D viscosity multiplied by the membrane thickness.  
The slight change of $D_{sup}^{-}$ in Fig.~\ref{Dphys2vsbeta} is the result of the change 
in the hydrodynamic screening length from $2/\nu$ to $1/\nu$. 

\begin{figure}[!ht]
\centering 
\includegraphics[width=0.5\textwidth]{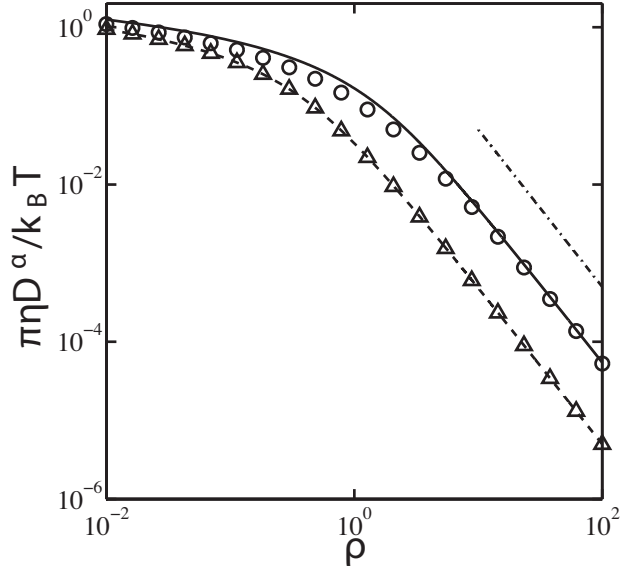} 
\caption{\label{Dphys2vsrho}$\pi \eta D^{\alpha}/k_{\rm B} T$ as a function of $\rho$ is shown for $ h=0.1$ and $\lambda=1$ for the supported case and 
in the limit of $\eta_f^+ \rightarrow 0$. $D^+$ and $D^-$ are shown by circular and triangular markers for the supported case, whereas they are shown by the solid and dashed lines in the limit of $\eta_f^+ \rightarrow 0$, respectively. $1/\rho^2$ is shown by a dash-dot line.} 
\end{figure}

The diffusion coefficient as a function of $\rho$ for a supported bilayer is shown in Fig.~\ref{Dphys2vsrho}. 
We note that the difference between $D^+$ and $D^-$ increases for both cases with increasing $\rho$ as shown in Fig.~\ref{Dphys2vsrho}. 
This implies that smaller sized domains have lower asymmetry between the upper and the lower leaflets, and it increases with increasing the size of the domain.

\section{Correlated dynamics between two leaflets}
\label{sec:CorrelatedDynamics}
So far, we have studied the diffusion of a liquid domain in either side of the leaflet of bilayer. 
According to the strength of the inter-leaflet coupling, 
the diffusion in a leaflet influences the diffusion in another leaflet. 
In this section, we study the correlated diffusion of two point particles in different leaflets. 

We define the $x$-axis using the line connecting the initial positions of two particles. 
By introducing the mobility tensor  
\cite{Ramachandran11,camleybrown2013}
\begin{equation}
G_{ij}^{\alpha \beta}({\bm k}) =M^{\alpha \beta} (k) \left( \delta_{i j} - \frac{k_i k_j}{k^2} \right), 
\end{equation}
where $M^{\alpha \beta} (k)$ is given in Eq.~(\ref{eq:green2}), 
we can calculate the correlation between displacements, 
$\left<\Delta r_i^{(1)}\Delta r_j^{(2)}\right>_r^{\alpha \beta}=2k_{\rm B} Tt G_{ij}^{\alpha \beta}(r)$, 
by the inverse Fourier transformation.

We consider the longitudinal and transverse coupling of displacements for two particles. 
The longitudinal coupling diffusion coefficient is defined by 
$D^{\alpha \beta}_L(r)\equiv \left<\Delta r_x^{(1)}\Delta r_x^{(2)}\right>_r^{\alpha \beta}/(2t)$ and 
is expressed as,
\begin{equation}
D^{\alpha \beta}_L(r) = k_{\rm B} T \int_0^\infty \frac{\text{d} k}{2 \pi} \frac{J_1(kr)}{r} M^{\alpha \beta} (k). 
\end{equation}
On the other hand, the transverse coupling diffusion coefficient is defined by 
$D^{\alpha \beta}_T(r)\equiv \left<\Delta r_y^{(1)}\Delta r_y^{(2)}\right>_r^{\alpha \beta}/(2t)$ and 
is expressed as, 
\begin{equation}
D^{\alpha \beta}_T(r) = k_{\rm B} T \int_0^\infty \frac{\text{d} k}{2 \pi} 
\frac{\partial J_1(kr)}{\partial r} M^{\alpha \beta} (k) ,  
\label{eq:transverse}
\end{equation}
where $\partial J_1(z)/\partial z=J_0(z)-J_1(z)/z$ should be noted~\cite{abram-stegun}.
Using appropriate function for $M^{\alpha\beta}(k)$, the above expressions can be evaluated for all the cases discussed in the previous sections. 

As a relevant example, we calculate the coupling displacements for the supported case mentioned in section \ref{sec:supportedbly}.
Using the appropriate mobility tensor, we obtain the longitudinal diffusion coefficient by using Mathematica as~\cite{mathematica}
\begin{equation}
D^{\alpha \beta}_L(r)= \frac{k_{\rm B} T}{4 \pi \eta \rho}\sum_{i=1}^4 \frac{h^{\alpha \beta}(\omega_i)}{f'(\omega_i)}\left[\frac{-2}{\omega_i\rho}+\pi\left\{\mathrm{Y}_1(-\omega_i\rho)+\mathbf{H}_{-1}(-\omega_i\rho)\right\}\right], 
\label{eq:corrlong}
\end{equation}
where $h^{\alpha \beta}(x)=g^{\alpha}(x)\delta_{\alpha \beta} + \lambda (1-\delta_{\alpha \beta})$, and $\omega_i$, 
$g^{\alpha}(x)$ and $f(x)$ were defined in Eq.~(\ref{Dphys}). 
$\mathrm{Y}_1(z)$ is the Bessel function of the second kind of order $1$, and 
$\mathbf{H}_{-1}(z)$ is the Struve H-function of order $-1$~\cite{abram-stegun}. 
Here, $\rho$ denotes the distance between two particles 
normalized by $\nu$, $\rho=\nu r$.
The transverse coupling diffusion coefficient   is calculated from Eq.~(\ref{eq:transverse}) by changing the order of integration and differentiation 
~\cite{mathematica}  
\begin{multline}
D^{\alpha \beta}_T(r) = \frac{k_{\rm B} T}{8 \pi \eta} \sum_{i=1}^4 \frac{\omega_i h^{\alpha \beta}(\omega_i)}{f'(\omega_i)}\left[\frac{2(2+\omega_i \rho)}{(\omega_i \rho)^2} -\pi\{\mathrm{Y}_0(-\omega_i \rho) - \mathrm{Y}_2(-\omega_i \rho) \right. \\\left. 
\vphantom{\frac{a}{b}}  + \mathbf{H}_{-2}(-\omega_i \rho) -\mathbf{H}_0(-\omega_i \rho)\}\right]. 
\label{eq:corrtran}
\end{multline}
Equations (\ref{eq:corrlong})-(\ref{eq:corrtran}) are obtained for point particles and valid without any restrictions on the separation distance. 
We study the above equations by taking the limits for $\rho$. 
Equations (\ref{eq:corrlong}) and (\ref{eq:corrtran}) become  
\begin{equation}
D^{\alpha \beta}_{L,T}(r) \approx \frac{k_{\rm B} T}{4 \pi \eta} \sum_{i=1}^4 \frac{\omega_i h^{\alpha \beta} (\omega_i)}{f'(\omega_i)}\left[\ln\left(\frac{-2}{\omega_i \rho}\right)-\gamma\pm \frac{1}{2}\right],   
\end{equation}
in the limit of $\omega_i \rho \ll 1$, and 
\begin{equation}
D^{\alpha \beta}_{L,T}(r) \approx \frac{k_{\rm B} T}{2 \pi \eta \rho^2} \sum_{i=1}^4 \left(\frac{\mp h^{\alpha \beta} (\omega_i)}{\omega_i f'(\omega_i)}\right) ,  
\end{equation}
in the limit of $\omega_i \rho \gg 1$. 
The upper and lower sign corresponds to the longitudinal and transverse component, respectively. 

\begin{figure}[!ht]
\centering 
\includegraphics[width=1\textwidth]{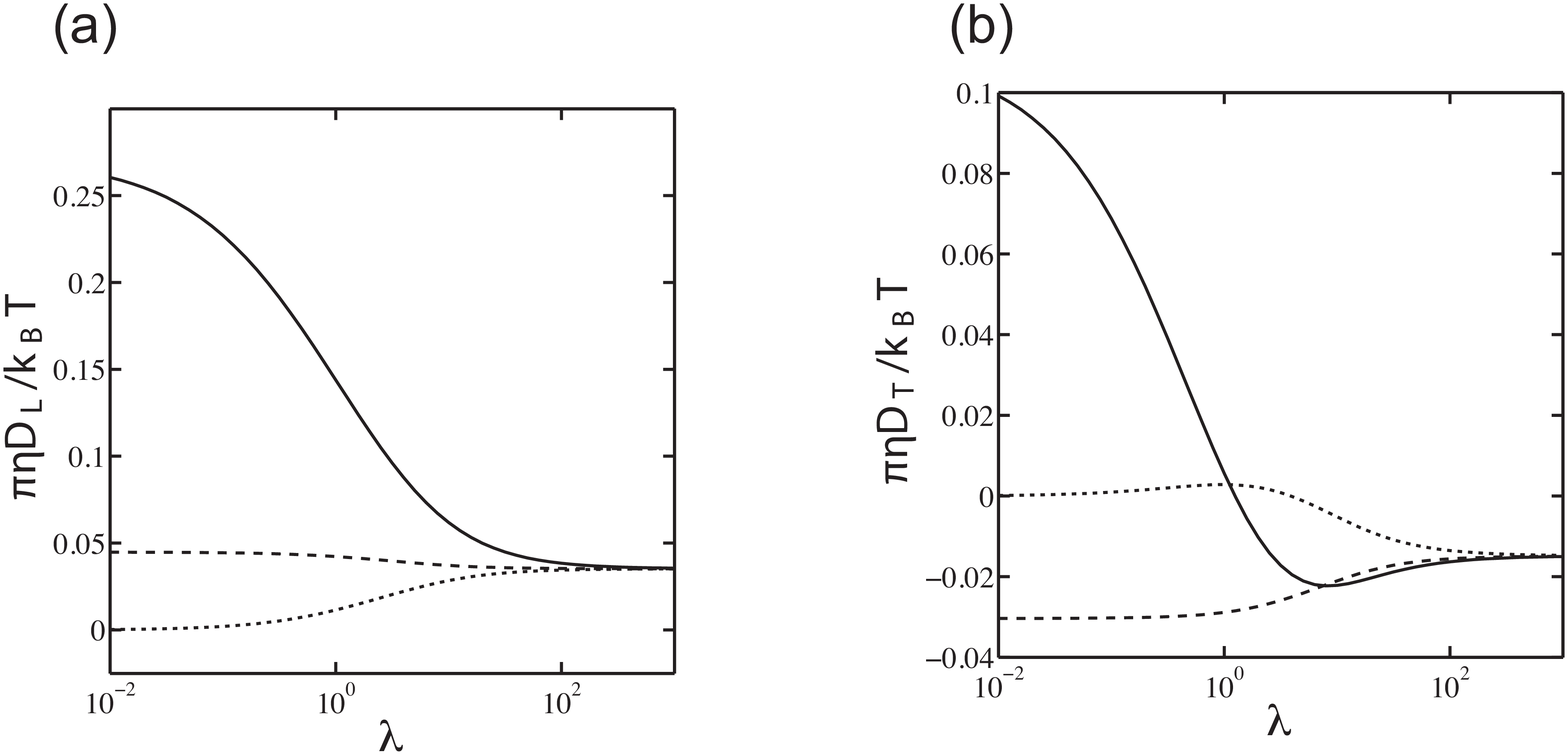} 
\caption{\label{fig:Corrlongbeta} The (a) longitudinal and (b) transverse correlation 
as a function of $\lambda$ calculated from Eq.~(\ref{eq:corrlong}) and Eq.~(\ref{eq:corrtran}), respectively. 
The solid line indicates the case that both particles are embedded in the upper leaflet, 
whereas the dashed line indicates the case that both particles are embedded in the lower leaflet. 
The dotted line indicates that two particles are embedded in different leaflets. Parameters used are $ h=0.1$, $\rho=1$.}
\end{figure}

\begin{figure}[!ht]
\centering 
\includegraphics[width=1\textwidth]{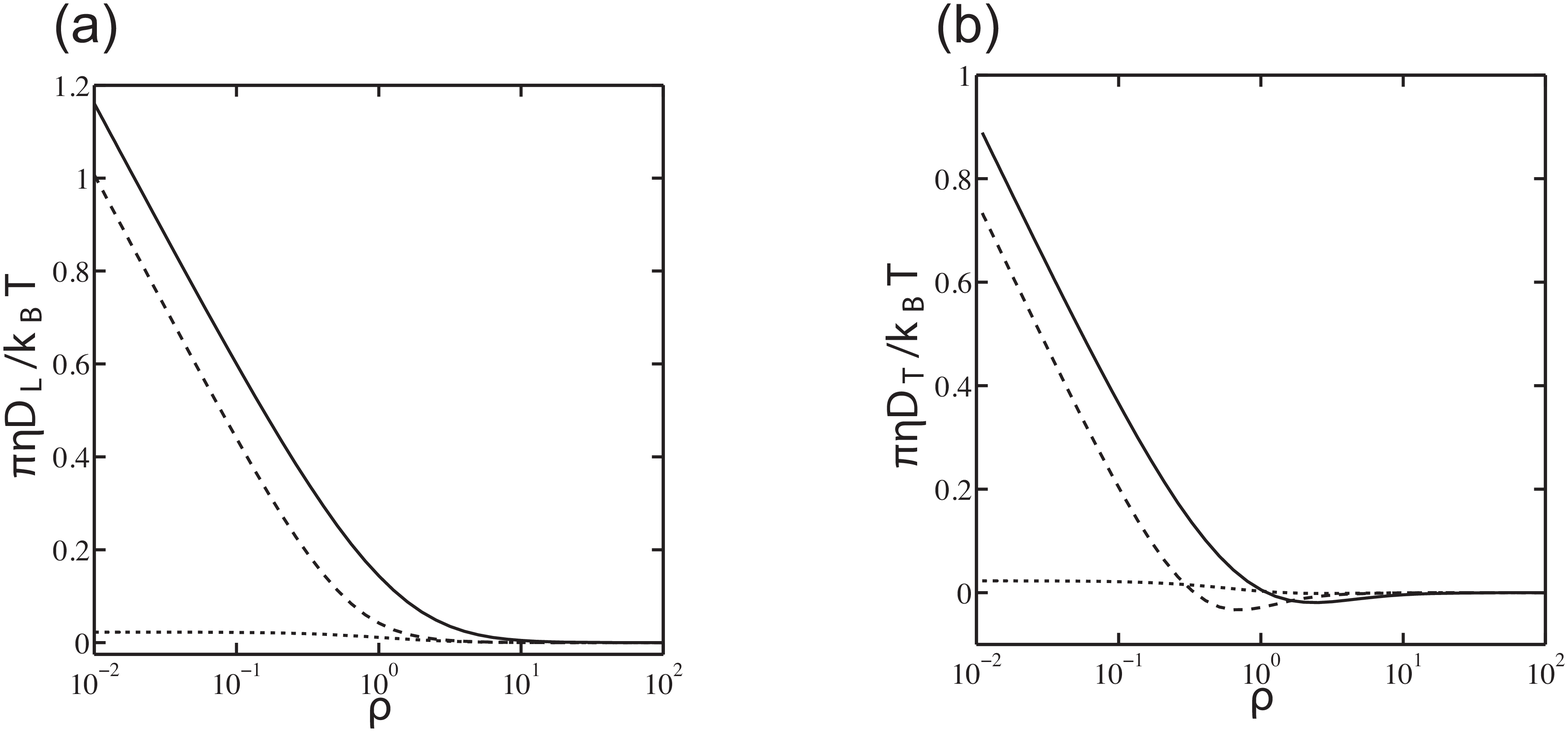} 
\caption{\label{fig:Corrlongrho}The (a) longitudinal and (b)  transverse correlation as a function of $\rho$ calculated from 
Eq.~(\ref{eq:corrlong}) and Eq.~(\ref{eq:corrtran}), respectively. 
The solid line indicates the case that both particles are embedded in the upper leaflet, 
whereas the dashed line indicates that both particles are embedded in the lower leaflet. The dotted line indicates that two particles are embedded in different leaflets. Parameters used are $ h=0.1$, $\lambda=1$.}
\end{figure}

The correlated diffusion coefficients as a function of the frictional coupling are shown in Fig.~\ref{fig:Corrlongbeta}. 
From Fig.~\ref{fig:Corrlongbeta} (a), 
we note that the longitudinal correlated diffusion exhibits a similar behavior as the diffusion coefficient   
if $\alpha=\beta$. 
As the value of the frictional coupling increases, 
the correlation decreases  
when the particles are in the same leaflet, 
whereas it increases when the particles are in the different leaflets. 
It is seen that the correlated diffusion between two particles in the different leaflets 
is small when $\lambda\leq 1$. 
However, as the value of the frictional coupling increases,  
the correlated diffusion coefficient in the different leaflets tends to converge to 
the smaller value of the correlated diffusion in the same leaflet.  

As shown in Fig.~\ref{fig:Corrlongbeta} (b), the transverse components show some interesting features. 
For particles in the same leaflet, the correlation reduces from a positive value to a negative value for particles in the 
upper leaflet on increasing inter-membrane friction, 
while the correlation remains negative and approaches the value of the upper leaflet 
when particles are in the lower leaflet. 

The correlated diffusion coefficients as a function of the separation between two inclusions are shown in Fig.~\ref{fig:Corrlongrho}. 
The longitudinal correlated diffusion exhibits a similar behavior as the diffusion coefficient. 
The transverse component decreases to a negative value and turns to increase as the separation increases if $\alpha=\alpha'$ .  
In the description of the bilayer as a uniform 2D fluid, 
a similar non-monotonic dependence of 
the coupling diffusion coefficients on the separation between
the inclusions has been reported \cite{Oppenheimer}.
The correlated diffusion coefficient in different leaflets monotonically decreases to zero by increasing the separation.


\section{Summary and Discussion}
\label{sec:Discussion}
We have investigated the effect of inter-leaflet friction on the domain diffusion in a leaflet of a bilayer. 
We considered both symmetric and asymmetric environments of solvents. 
For both cases, we have shown that the effect of the inter-leaflet friction on the leaflet diffusion is important if 
$\lambda$ is smaller than $\lambda^*_2=1/(2 h)$;  
the value is taken from the inflection point of diffusion coefficient as a function of the inter-leaflet friction for large domains  
(see Appendix C for details).  
For small domains, 
the inflection point can be characterized by $
\lambda^*_1=(1/\rho-\cot( h/\rho))/(2\rho)$.  
The critical domain size can be obtained from $\lambda^*_1=\lambda^*_2$. 
We have also studied the size dependence of the leaflet-diffusion coefficient.

According to recent experiments, 
the value of inter-leaflet friction is estimated to be $\Lambda \sim 10^8-10^9$ Pa$\cdot$s/m~\cite{Bitbol,Evans94,Fournier,Mechan}. 
The relation, $\lambda^*_2=1/(2 h)$, can be rewritten as $\Lambda^*=\eta_f/(2H)$. 
By substituting the typical values, $\eta_f \sim 1 \times 10^{-3}$ Pa$\cdot$s and 
$H \sim 1$ nm, 
we find $\Lambda^* \approx 5 \times 10^5$ Pa$\cdot$s/m. 
However, if solvents are confined in molecular scales, 
the solvent viscosity is reported to be significantly increased, $\eta_f \sim 0.1$ Pa$\cdot$s~\cite{camleybrown2013,Zhang}.
In the extremely confined situation, 
$\Lambda^*$ can be increased to $5 \times 10^7$ Pa$\cdot$s/m. 
Experimental values of $\Lambda \sim 10^8-10^9$ Pa$\cdot$s/m 
is not much different from the value of $\Lambda^*$ for the extremely confined case. 
The value of $\lambda^*_1$ relevant to the small size of diffusing 
objects can be order of magnitude larger than that of $\lambda^*_2$. 
For small domains, by using $\lambda^*_1$ we estimate that 
$\Lambda^*$ can be $10^9$ Pa$\cdot$s/m for the extremely confined case and 
$10^7$ Pa$\cdot$s/m without molecular scale confinements. 

According to simulations, 
the value of inter-leaflet friction is reported to be $\Lambda \sim 2.4 \times 10^6$ Pa$\cdot$s/m~\cite{Shkulipa,denOtter}. 
There, 
the leaflet diffusion at small sizes can be affected by 
the inter-leaflet friction. 
Indeed, the lateral diffusion 
of the lower leaflet was shown to be slower than that of the upper leaflet by the
attractive interaction of the support 
within a coarse-grained molecular simulation~\cite{Xing}.

In this paper, we have studied 
asymmetry in leaflet diffusion coefficients induced by the asymmetry in environments. 
By anchoring lipopolymer, the fluidity of a leaflet composing a bilayer can be obstructed and 
the fluidity of leaflets can be asymmetric~\cite{Tanaka05,HYZhang}. 
When the fluidity of leaflets is asymmetric, the inter-leaflet friction should be in principle taken into account. 
Although the model studied in this paper is simpler than the situation of lipopolymer-grafted
bilayers, 
our study could be useful to understand 
the effect of the asymmetry in the fluidity between two leaflets 
on the leaflet diffusion. 

Finally, 
we discuss the diffusion of a circular liquid domain in an isolated membrane without any solvent 
to study the actual contribution of the momentum dissipation by the inter-leaflet friction. 
In the absence of the coupling to 3D solvent,  
the SD length given by the ratio
between the 2D membrane viscosity and the 3D solvent viscosity diverges. 
This divergence reflects the so-called Stokes paradox; 
the diffusion coefficient   cannot be obtained 
by solving hydrodynamic equation in the limit of low-Reynolds number in pure 2D fluid 
since the influence of translational motion of a circular body 
extends to large distances due to lack of enough momentum dissipation in 2D. 
Although the coupling between solvents and the bilayer is absent, 
the momentum can be dissipated through the inter-leaflet friction. 
In order to see whether the Stokes paradox can be resolved by inter-leaflet friction alone, 
we set  $\eta_f^\pm=0$ in Eq.~(\ref{eq:dgen}) 
resulting,
\begin{equation}
D=\frac{k_{\rm B} T}{\pi R^2}\int_0^\infty {\rm d}k 
\frac{\left[J_1 (kR)\right]^2}{k}\left[\frac{\eta_- k^2 +b}{k^2\left(\eta^+ \eta^- k^2 + \Lambda \eta^+\eta^-\right)}\right] .
\label{eq:nofluid}
\end{equation}
The above expression can be rewritten as 
\begin{equation}
D=\frac{k_{\rm B} T}{\pi R^2} \frac{1}{\eta^+ + \eta^-} 
\left\{ \int_0^\infty {\rm d}k \frac{\left[J_1 (kR)\right]^2}{k^3} + \frac{\eta^-}{\eta^+}
\int_0^\infty {\rm d}k \frac{\left[J_1 (kR)\right]^2}{k} \left[\frac{1}{k^2 + \Lambda (\eta^+ + \eta^-)/(\eta^+ \eta^-)}\right]\right\} . 
\label{eq:dnofluid}
\end{equation}
In Eq.~(\ref{eq:dnofluid}), we note that the first term diverges at $k=0$, because $J_1(z)\approx z$ when $z \ll 1$ and $J_1(kR)^2/k^3 \approx 1/k$ in this limit. 
This shows that frictional coupling of leaflets is not enough to provide for a finite diffusion coefficient, and contribution from solvents is essential.
We get an interesting result if we allow $\eta^- \to \infty$, which implies that the lower leaflet of the membrane behaves as a solid. 
In this limit, we obtain
\begin{equation}
D=\frac{k_{\rm B} T}{\pi R^2}\int_0^\infty {\rm d}k \frac{\left[J_1 (kR)\right]^2}{k} \left[\frac{1}{\eta^+ k^2 + \Lambda }\right] .
\label{eq:nofluidlimit}
\end{equation}
Equation (\ref{eq:nofluidlimit}) does give a finite result for the diffusion coefficient, which can be explained by the lower leaflet providing required friction for diffusion, instead of the solvent.

In the above, we show that 
the frictional coupling between two leaflets alone is not enough to resolve the Stokes paradox. 
Only in the case that one of the leaflet of the membrane is immobile, 
a finite diffusion coefficient can be obtained without solvents. 
The particular model was already applied to study the inter-leaflet friction from the diffusion in a leaflet coupled to  
an immobile leaflet \cite{Merkel},  
and the diffusion in a polymer-supported monolayer 
where the polymer plays the roll of an immobile leaflet \cite{Tanaka07}.
However, if both leaflets are viscous, 
the model should include the coupling to solvents to have finite diffusion coefficients.

\acknowledgments
KS and SK are supported by Grant-in-Aid for Scientific Research
(grant No.\ 24540439) from the MEXT of Japan.
This research was conducted under Agreement for Collaboration
between IIT Bombay and Nanosystem Research Institute, AIST. 
SK would like to acknowledge support from the Grant-in-Aid for Scientific
 Research on Innovative Areas "Fluctuation \& Structure" (No. 25103010)
 from the MEXT of Japan, and the JSPS Core-to-Core Program "International
 research network for nonequilibrium dynamics of soft matter".

\appendix 
\section{{Velocity field}}
Denoting  the total force on the circular domain in the steady state by ${\bm F}^\pm$, 
the velocity field is expressed as \cite{note1}
\begin{equation}
{\bm v}^\alpha({\bm r})=- \left[\int_0^\infty{\rm d}k\, M^{\alpha \alpha} (k) \frac{J_1(kR)J_1(kr)}{\pi kRr} \,\hat{\bm e}_x +
\int_0^\infty{\rm d}k\, M^{\alpha \alpha} (k) \frac{J_1(kR)J_2(kr)\sin(\theta)}{\pi R} \,\hat{\bm e}_{\theta}\right]F^\alpha , 
\label{eq:vfields}
\end{equation}
where $\hat{\bm e}_x$ is the unit vector in the direction of the $x$-axis, 
$\hat{\bm e}_{\theta}$ is the unit vector tangential to the circle pointed in the direction of rotation, indices $\alpha$, $\beta$ denote the leaflets. 
In the above equation, 
$M^{\alpha \beta}( k )$ represents a component of a mobility tensor in Fourier space and is given by \cite{camleybrown2013}
\begin{equation}
M^{\alpha \beta}( k )=\frac{[\eta^{(-\alpha)}k^2+\eta_f^{(-\alpha)}k\coth(kH^{(-\alpha)})]\delta_{\alpha\beta}+\Lambda}
{\prod_{\alpha} (\eta^{(\alpha)}k^2+\eta_f^{(\alpha)}k\coth(kH^{(\alpha)})+\Lambda)-\Lambda^2} ,
\label{eq:green2}
\end{equation}
where $(-\alpha)$ represents the leaflet opposite to $\alpha$. 

By noticing that the velocity at $(R,0)$ is equal to ${\bm U}$,  
the friction coefficient, $\zeta$, is obtained from the linear relation,  
${\bm U} = -(1/\zeta) {\bm F}$ . 
Using the Einstein relation $D=k_{\rm B}T /\zeta$ \cite{Kubobook}, we obtain the diffusion coefficient as in Eq.~(\ref{eq:dgen}).

\begin{figure}[!ht]
\centering 
\includegraphics[width=\textwidth]{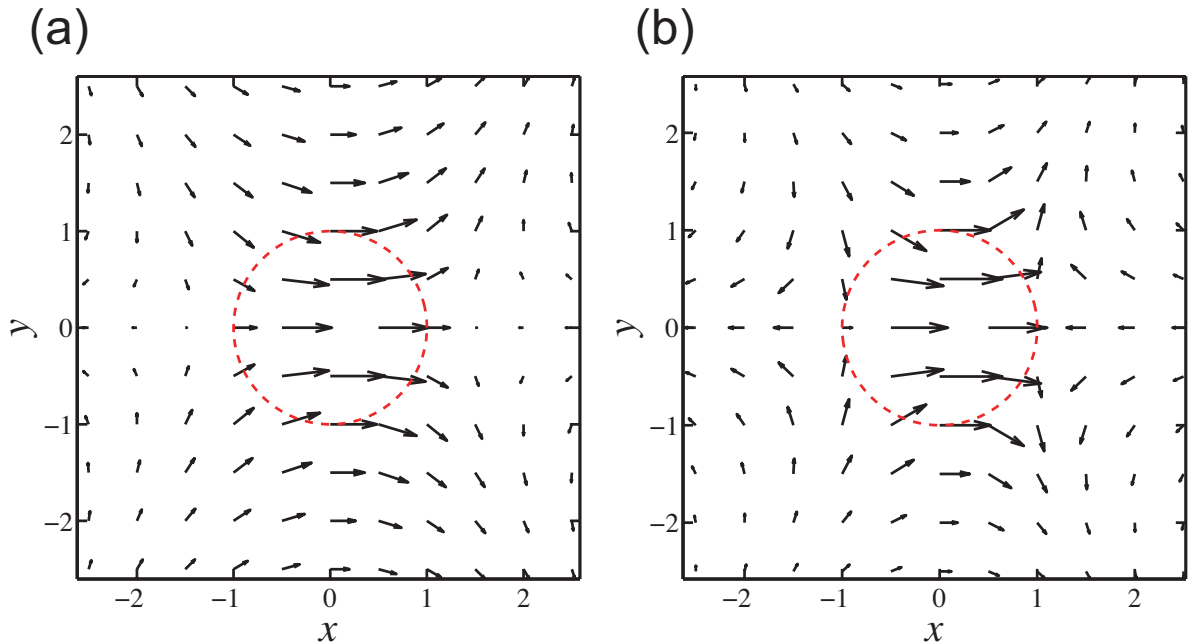} 
\caption{\label{fig:vfieldphysp} Velocity field for the supported case calculated from Eq.~(\ref{Dphys}) when the domain is in 
the (a) upper  and (b) lower  leaflets. 
The center of the domain moves with the velocity $\bm U$ in the positive direction of the $x$-axis. Parameters used are $ h=0.1$, $\rho=1$, $\lambda=1$.}
\end{figure}

The velocity fields can be drawn by using Eq.~(\ref{eq:vfields}). 
As an example, 
we draw the velocity fields for the supported case of Sec.~\ref{sec:supportedbly}. 
The velocity field in the lower leaflet is shown when a liquid domain is located in the upper (lower) leaflet in Fig.~\ref{fig:vfieldphysp}. 
By comparing the velocity fields, we could see that a part of the velocity field in the lower leaflet flows opposite to the flow 
velocity of the liquid domain compared to that in the upper leaflet. 

\section{{Partial Fraction expansion to solve required integrals}}
We now derive the result in Eq.~(\ref{eq:pfsym}) from Eqs.~(\ref{eq:drel}) and (\ref{eq:dsym}) using the method of partial fractions. 
A general relation used to obtain partial fraction expansion for $g(x)/f(x)$ is \cite{Arfken}
\begin{equation}
\frac{g(x)}{f(x)}=\sum_{\text{all roots}}\frac{g(x_i)}{f'(x_i)}\frac{1}{(x-x_i)},
\label{eq:pfexp}
\end{equation}
where $x_i$ are the roots of the denominator, $f(x)$, and the sum is carried out over all roots.
We find the partial fraction expansion for the $D_1(\lambda)$ term in Eq.~(\ref{eq:drel}). 
The other term can be obtained by putting $\lambda=0$ and following the same procedure. 
The fraction under consideration is $1/(\kappa^2+\kappa \coth(\kappa h)+ 2\lambda)$. 
Equating the denominator to zero, we obtain,
\begin{equation}
\kappa_s^2+ \kappa_s \coth(\kappa_s  h)+ 2\lambda = 0 ,
\end{equation}
where `$s$' is the index of the root. 
It is convenient to introduce a transformation $\kappa_s = i \omega_s$ when an infinite number of roots exists. 
This gives,
\begin{equation}
\omega_s^2=\omega_s \text{cot}(\omega_s  h) + 2\lambda. 
\label{eq:muexp}
\end{equation}
We substitute $\text{cot}(\omega_s  h)=(\omega_s^2-2\lambda)/\omega_s$ in Eq.~(\ref{eq:pfexp}) to obtain,
\begin{equation}
\frac{1}{\kappa^2+\kappa\coth(\kappa  h) + 2\lambda}=\sum_{s=-\infty}^\infty \left(\frac{1}{ h\omega_s^4+(1+ h-4\lambda h)\omega_s^2+2\lambda+4 h\lambda^2}\right)\frac{\omega_s/i}{\kappa-i \omega_s} ,
\end{equation}
where we can assume $\omega_{-s}=-\omega_s$ without loss of generality. 
Since the expression within the parenthesis is even, we obtain
\begin{equation}
\frac{1}{\kappa^2+ \kappa\coth(\kappa h) + 2\lambda}=\sum_{s=1}^\infty \left(\frac{2\omega_s^2}{ h\omega_s^4+(1+ h-4\lambda h)\omega_s^2+2\lambda+4 h\lambda^2}\right)\frac{1}{\kappa^2+ \omega_s^2} . 
\end{equation}
If the first term in the expansion dominates, Eq.~(\ref{eq:dsym}) becomes 
\begin{equation}
D_1 (\lambda)=\frac{k_{\rm B} T}{\pi \eta \rho^2}\,\frac{2}{ h\omega_1^4+(1+ h-4\lambda h)\omega_1
^2+2\lambda+4 h\lambda^2}\left[ \frac{1}{2}-I_1( \omega_1\rho) K_1(\omega_1\rho)\right] . 
\label{eq:dbetapfexp}
\end{equation}

\section{{Inflection point of the diffusion coefficient as a function of $\lambda$}}
The inflection region for the diffusion coefficient as a function of $\lambda$ is 
very different when $\rho \ll 1$  in Fig.~\ref{DsymvsBeta}.
The inflection originates from the variation of the Bessel functions. 
When $\omega_1\rho \ll 1$, the term including Bessel functions can be approximated by a logarithmic function of the variable by using 
the expansion, $I_1(z) K_1(z)/z^2 \sim (1/4) \ln z$ for $z\leq 1$. 
The logarithmic dependence is weak and the condition, $\omega_1\rho \gg 1$, should be satisfied 
to obtain sharp variable dependence from the Bessel functions. 
By substituting $\omega_1\rho \sim 1$  into the characteristic equation, Eq.~(\ref{eq:characteristiceq}), we obtain  
$\lambda^*_1=[1/\rho-\cot( h/\rho)]/(2\rho)$. 
At the inflection point, 
the term including Bessel functions should depend sharply on the variables and 
the variable itself should depend sharply on the value of $\lambda$.  

We now discuss the $\lambda$-dependence of the variable $\omega_1\rho$. 
In the limit of $ h \rightarrow 0$, we have $\cot( h/\rho) \sim \rho/ h$ and 
obtain, $\omega_1 \rho \sim\left( \rho/\sqrt{ h} \right) \sqrt{1+ 2 \lambda  h}$. 
Therefore, the variable,  $\omega_1 \rho$, depends largely on $\lambda$  if $\lambda$ value exceeds $\lambda^*_2=1/(2  h)$. 
By comparison, we note that 
$\lambda^*_2$ can be used even when $ h$ is as large as $10$. 
In summary, the larger of $\lambda^*_1$ or $\lambda^*_2$ characterizes the inflection point. 
When $\rho$ is small, the inflection point is characterized by $\lambda^*_1$. 
By increasing $\rho$, $\lambda^*_1$ decreases and the inflection point is characterized by $\lambda^*_2$. 
The switching between $\lambda^*_1$ and $\lambda^*_2$ takes place at $\rho$ satisfying $\lambda^*_1=\lambda^*_2$.
As seen in Fig. \ref{DsymvsBeta}, if the inflection point is characterized by $\lambda^*_1$, 
the inflection point can become order of magnitude larger than $\lambda^*_2$ by decreasing $\rho$. 

The value characterizing the inflection point, $\lambda^*_2$, can be obtained by equating 
the inter-leaflet sliding length, $\sqrt{\eta/b}$, with the ES hydrodynamic screening length, $\sqrt{ h}/\nu$.
The velocity field in the membrane perturbed by a diffusing particle is 2D-like within 
the  ES length and becomes 3D-like 
if the distance from the diffusing particle exceeds the ES length.
If the inter-leaflet sliding length is smaller than the  ES length, 
 2D-like feature of the velocity field is not 
affected by inter-leaflet sliding. 
  On the other hand, if the inter-leaflet sliding length is larger than the  ES length, 
 2D-like feature of the velocity field is largely influenced by the inter-leaflet sliding and 
the diffusion coefficient is increased by the sliding effect.



\begin{references}
\bibitem{note1}
 This result is the correction of a similar result from a previous publication \cite{sekiramakomura2011}.
The velocity at $(R,0)$ was correct  and the other parts are not affected 
by the correction in \cite{sekiramakomura2011}.
\end{references}
\end{document}